\documentclass[12pt]{JHEP}
\usepackage{graphicx,amsmath,amssymb}
\usepackage{epsfig,multicol}
\usepackage{epsf}
\usepackage{epstopdf}
\input{epsf.sty}

\newcommand{\bas}{\begin{eqnarray*}}
\newcommand{\eas}{\end{eqnarray*}}

\newcommand{\ba}{\begin{eqnarray}}
\newcommand{\ea}{\end{eqnarray}}

\newcommand{\oo}{\Omega}
\newcommand{\e}{\epsilon}

\newcommand{\la}{\lambda}

\title{Lumps in the throat}
\author{Keshav Dasgupta\footnote{keshav@hep.physics.mcgill.ca}~,
Hassan Firouzjahi\footnote{firouz@hep.physics.mcgill.ca}~ and ~ Rhiannon Gwyn\footnote{gwynr@hep.physics.mcgill.ca} \\
\noindent Rutherford Physics Building, McGill University, Montreal, Canada, H3A 2T8}

\abstract{We study classical lump solutions in a warped throat where brane inflation takes place.
Some of the solitonic or lump solutions that 
we study here are the $(p,q)$ cosmic strings and their junctions, cosmic necklaces and semi-local strings and generic 
semi-local defects. We show how various wrapping modes of D3-branes may be used to study all these 
defects in one interpolating set-up.
Our construction allows us to study $(p,q)$-string junctions in curved backgrounds and in the 
presence of non-trivial RR fluxes. 
We extend the junction construction to allow for the possibility of cosmic necklaces, and show how these new lump
solutions form a consistent picture in the inflationary brane models. We also give a generic construction of semi-local 
defects in these backgrounds, and argue that our construction encompasses all possible constructions of semi-local 
defects with any global symmetries. The cosmological implications of these configurations are briefly studied.
} 
\keywords{Warped geometry, D-branes, Cosmic strings}

\preprint{}

\begin{document}

\section{Introduction}
Recently there has been a renewal of interest in cosmic strings \cite{Kibble:2004hq}. This is due to the realization that cosmic strings
can be formed \cite{Sarangi:2002yt, Jones:2003da}
in brane inflation scenarios
\cite{Dvali:1998pa, Burgess:2001fx, Dvali:2001fw}. Similarly in some brane inflation models 
semi-local strings \cite{Vachaspati:1991dz}
are also produced \cite{Urrestilla:2004eh, Binetruy:2004hh, Dasgupta:2004dw}. 
Detecting a cosmic or semi-local string could potentially open a unique window of opportunity 
into string theory. One may obtain important information about the parameters of the theory by looking at the details of the spectrum of these strings.

At the end of brane inflation, both fundamental strings, i.e. F-Strings, and D1-branes, i.e.
D-strings, are formed. From F- and D-strings, one can construct the $(p,q)$-string which is a bound state of $p$ F-strings and $q$ D-strings. 
When different types of cosmic superstrings intersect they cannot intercommute; instead they form junctions. This can have important consequences for the evolution of cosmic superstring networks \cite{Jackson:2004zg, Tye:2005fn, Copeland:2006if}.  
In more realistic models of brane inflation in warped compactification, such as in \cite{Kachru:2003sx}, the spectrum of $(p,q)$-strings can be  non-trivial \cite{Firouzjahi:2006vp, Firouzjahi:2006xa}.
%The evolution of these $(p,q)$ cosmic superstring  was studied recently in 
%\cite{Leblond:2007tf}. It was argued that the system reaches a scaling regime.

%In addition to that, 
Sometimes
in the presence of enough global symmetries in brane inflation models that involve both D3- and D7- branes,
\cite{Dasgupta:2002ew, Hsu:2003cy, Chen:2005ae}, 
semi-local defects or semi-local
strings can be formed instead of the stable cosmic strings. These strings satisfy the current CMB constraints and appear to resolve 
all the issues that a copious generation of cosmic strings might produce. It is also known that formation of 
semi-local strings does not pose any cosmological problems \cite{Achucarro:1998ux}. In fact the 
formation rate of semi-local strings at weak gauge coupling is generically almost one third that of cosmic strings (and one 
quarter in two dimensions) \cite{Achucarro:1998ux}. 

In this paper we provide microscopic descriptions of different solitonic configurations in some warped backgrounds 
where brane inflation is supported. It turns out that all these solitons can be studied from various wrapping modes of 
D3-branes. For example, when a D3-brane with appropriate worldvolume electric and magnetic fluxes 
wraps a two-cycle in the Klebanov-Strassler (KS) type geometry \cite{Klebanov:2000hb}, this appears as
a ($p,q$)-string in our spacetime. A combination of three such D3-branes can be used succinctly to construct a three-string
junction in the throat region of the KS geometry. 
In section 2 we dwell on this issue, and show how the construction of 
three-string junctions of $(p,q)$-strings in the KS geometry can be made consistent with the modified tension formula in the 
throat. In addition to that, 
the understanding of string junctions is crucial to the evolution of cosmic superstring networks. 
%In the last part of this section we also study the supersymmetric properties of this junction.
In section 3 we present the microscopic interpretation of strings ending
on charged particles, like dyons or monopoles. These dyons or monopoles can be understood from our D3-branes wrapping 
the full three-cycle in the throat. These wrapped D3-branes lie at the intersection points of the 
three-string junctions and are called cosmic necklaces. In this section we also give some cosmological consequences 
of these necklaces. 

The next interesting construction is when the D3-branes wrap zero-cycles in the throat. In the presence of 
extra seven branes, the worldvolume theory on the D3-branes can support semi-local defects. The construction 
of these defects is subtle, and in section 4 we point out various conditions that need to be met so that 
semi-local defects can form. We also show that formation of these defects requires us to know the global behavior
of the background, at least from the end of the throat to the top. A full construction involves not only
knowledge of the global geometry, but also the presence of earlier ingredients like the junctions and cosmic strings.   
We present the first generic construction of these defects for an arbitrary choice of global symmetries. 

Finally in section 5 we discuss cosmological consequences of these solitonic solutions and we conclude in section 6.

\section{Three-string junctions in the throat}

The first of the three cases we will study involves the D3-branes wrapping two-cycles in the throat. These 
D3 branes do not slip out of the three-cycle because of the stabilising RR background fluxes that, via the 
underlying Myers effect \cite{Myers:1999ps}, are responsible for the existence of stable defects in our spacetime \cite{Thomas:2006ud, Firouzjahi:2006xa}. 
A more detailed construction to show how wrapped D-branes do not shrink to a zero-cycle when wrapped on non-topological
cycles is given in \cite{Bachas:2000ik}. 

\subsection{ Set-up}
In  brane-antibrane inflation both F- and D-strings are produced at the end of inflation \cite{Sarangi:2002yt, Jones:2003da, Copeland:2003bj, Leblond:2004uc}. These F- and D-strings can create $(p,q)$-strings, which are  bound states of $p$ F-strings and $q$ D-strings.
We are interested in $(p,q)$-string junctions in the Klebanov-Strassler background where brane inflation takes place, as in \cite{Kachru:2003sx}. 
The three-string junction in a flat background was studied in
\cite{Dasgupta:1997pu, Rey:1997sp, Sen:1997xi}.

The metric of the deformed conifold is studied in \cite{Candelas:1989js}.
At the tip of the deformed conifold the space ends smoothly at a round three-sphere with radius $\sim$ $\epsilon^{1/3}$. The metric at the tip is given by 
\ba
\label{metric0}
ds^2 \sim  h^2\, \eta_{\mu \nu} dx^{\mu} dx^{\nu}+
  g_s M \alpha'(d\psi^2 +\sin^2 \psi\, d \oo_2 ^2),
\ea
where $h$ is the warp factor at the bottom of the throat
\ba
h= \e^{2/3}2^{-1/6} a_0^{-1/4} (g_s M \alpha')^{-1/2},
\ea
 and  $a_0 \sim 0.72$ is a numerical constant in the KS solution\footnote {Actually in the KS solution
 there is another numerical constant, $b \sim 0.93$, and $g_s M \rightarrow g_s M b$ in the metric.
 For simplicity we set  $b=1$.} and $g_s$ is the perturbative string coupling.
 Here $\psi$ is the usual polar coordinate in an $S^3$, ranging from $0$ to $\pi$, and 
M represents the number of Ramond-Ramond fluxes $F_3$ turned on inside this $S^3$.
The two-form associated with the flux $F_3$ is given by
\ba
\label{C2}
C_{(2)}=  M \alpha' \,  \left(\psi-\frac{\sin (2\psi)}{2} \right) \sin \theta\, d \theta \, d\phi \, .
\ea
In the KS solution the axion $C_0=0$ and at the bottom of the throat $B_{ab}=0$.
On the gauge theory side $M$ corresponds to the 
$SU(M)$ gauge theory living at the tip of the conifold.

In \cite{Firouzjahi:2006vp} (see also \cite{Herzog:2001fq})
the $(p,q)$-string was interpreted as a wrapped D3-brane with $p$ units of electric flux and $q$ units of
magnetic flux. In this picture the D3-brane is wrapped around an $S^2$ inside the $S^3$ at the bottom 
of the throat. The $S^2$ is parametrized by the angle $\psi$.
As illustration, consider a D3-brane extended along the $X^0$ and $X^1$ directions, and wrapped 
around the $\theta$ and $ \phi$ directions at the bottom of the KS throat.
Effectively, this looks like a 1-dimensional object extended along  $X^0$ and $X^1$ directions.

The action of the D3-brane is
\ba
\label{action1}
S= \int d^4x\, {\cal L}=
 \int d^4x\, \left( -\mu_3 g_s^{-1} \sqrt{ -|g_{ab} +\la F_{ab} | } + \mu_3 \lambda C_{(2)}\, \wedge  F 
%+ p A_0\,\delta(x^1) \,\delta(\Omega_2)
\right).
\ea
Here $\mu_3$ is the D3-brane charge, $\la= 2\pi \alpha'$, and $F$ is the worldvolume gauge field. 
To accommodate the $(p,q)$-string in this picture, one requires $F_{\theta \phi}=\frac{q}{2} \sin\, \theta$ and 
$ 4\pi \delta {\cal L}/ \delta F_{01}=p$.

After integrating over the $\theta$ and $\phi$ directions, this leads to
\ba
\label{action2}
S= \int dt\, dx^1\, \left(-\Delta \sqrt{ h^4 - \la^2 F_{01}^2} +
\Omega\, F_{01} 
\right),
\ea
where $\Omega$ and $\Delta$ are defined by \cite{Firouzjahi:2006vp,Firouzjahi:2006xa}
:

\ba
\label{Omega}
\Omega \equiv \la \mu_3 \int_{S^2} C_{(2)} = \frac{M}{\pi}\, (\psi - \frac{1}{2} \sin\, 2\psi  )
\ea
and
\ba
\label{delta}
\Delta &\equiv& \mu_3 g_s^{-1} \int_{S^2} \sqrt{g_{\theta \theta} g_{\phi \phi} + \la^2 F_{\theta \phi}^2 }\nonumber\\
&=&\la^{-1} \sqrt{\frac{ M^2}{\pi^2}\, \sin^4 \psi +\frac{q^2}{g_s^2} }.
\ea
Effectively the action (\ref{action2}) represents a $(p,q)$ cosmic string extended along the $X^0$ and $X^1$ directions. The D-string charge is encoded in $\Delta$, while
the electric charge $p$ is represented by the $U(1)$ gauge field $A$ on the D-string worldvolume.
The Hamiltonian or tension of the cosmic string is
\ba
\label{H}
{\cal H}= \frac{h^2}{\la}\, \sqrt{ \frac{q^2}{g_s^2} + \frac{ M^2}{\pi^2}\, \sin^4 \psi 
+ \left( p - \frac{M}{\pi} (\psi -\frac{\sin 2\psi}{2}) \right)^2 } \, .
\ea

The stable configuration is given by 
\ba
\label{psi}
\psi= \frac{\pi\, p}{M},
\ea
 with the Hamiltonian
\ba
\label{E}
{\cal H}= \frac{h^2}{\la}\, \sqrt{ \frac{q^2}{g_s^2} + \frac{ M^2}{\pi^2}\,
 \sin^2 \left(  \frac{\pi p}{M}\right) }.
\ea
Interestingly, the F-strings are charged in $Z_M$ and are non-BPS. When $p=M$, the tension of the F-strings vanishes. 
From the dual gauge theory side the F-strings correspond to flux tubes connecting quarks and anti-quarks in $SU(M)$ gauge theory at the bottom of the throat. When $p=M$, the quarks and anti-quarks form baryons and anti-baryons respectively and the flux tubes disappear.

In the limit $M\rightarrow \infty$,  the bottom of the throat approximates a flat background and the
above expression reduces to the known formula for the $(p,q)$-string spectrum in a flat background:
\ba
\label{flat}
T=  \frac{1}{\la}\, \sqrt{ \frac{q^2}{g_s^2} +p^2 }.
\ea

\begin{figure}[t]
\vspace{-1cm}
  % \centering
   \hspace{-1.6cm}
   \includegraphics[width=3in]{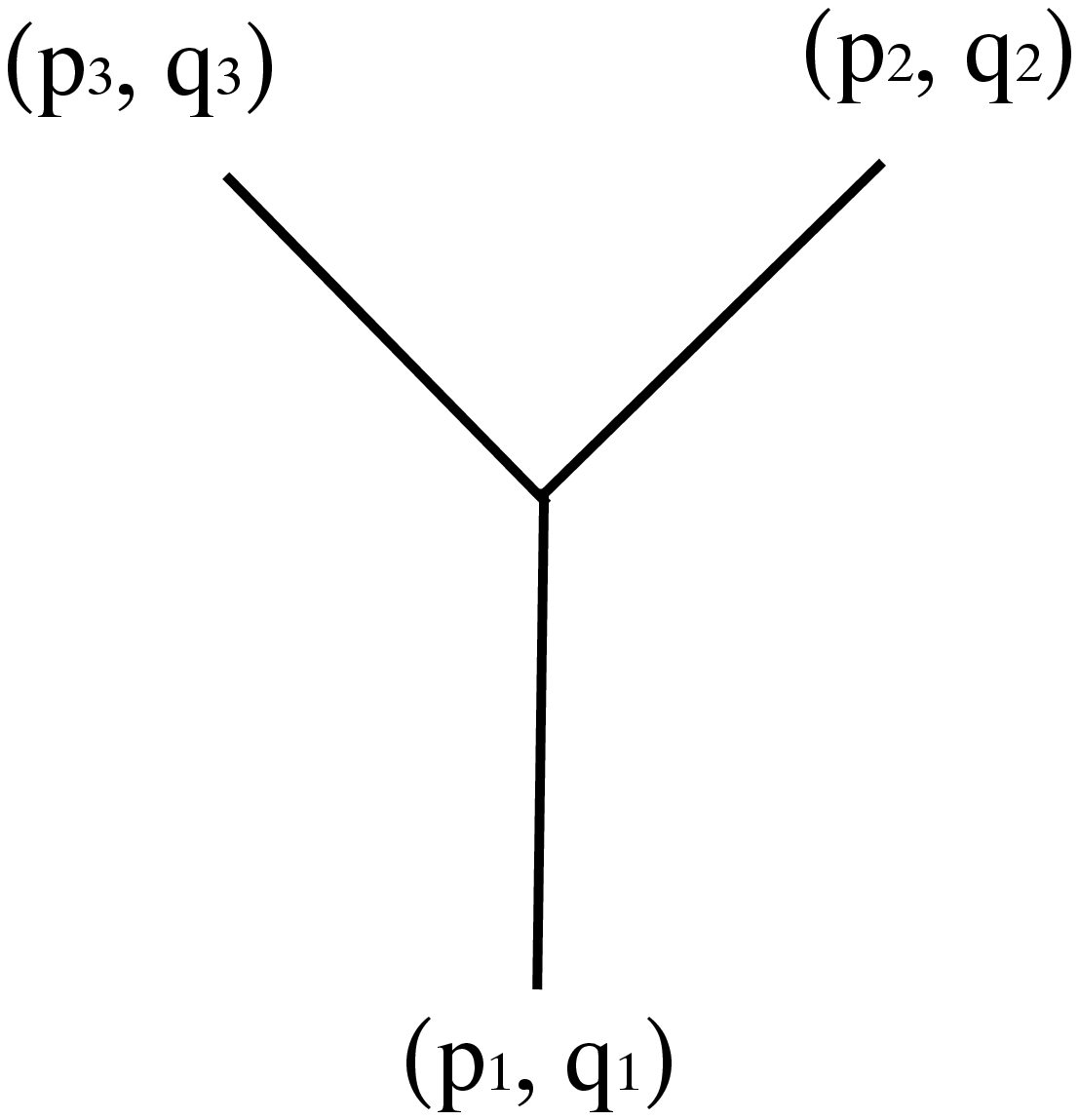} \hspace{-4.8cm}
   \includegraphics[width=2.8in]{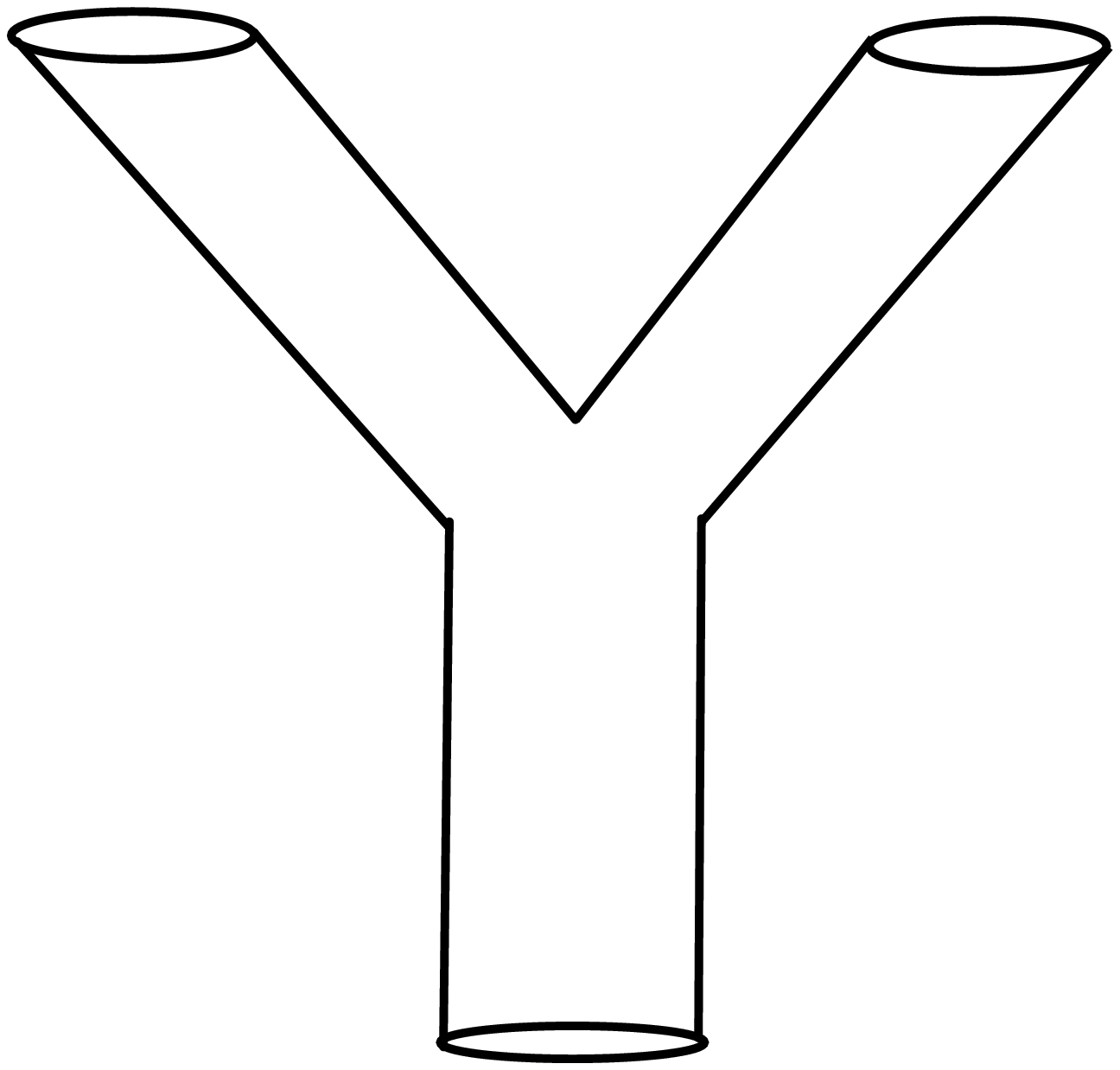}\hspace{-2cm}\includegraphics[width=3in]{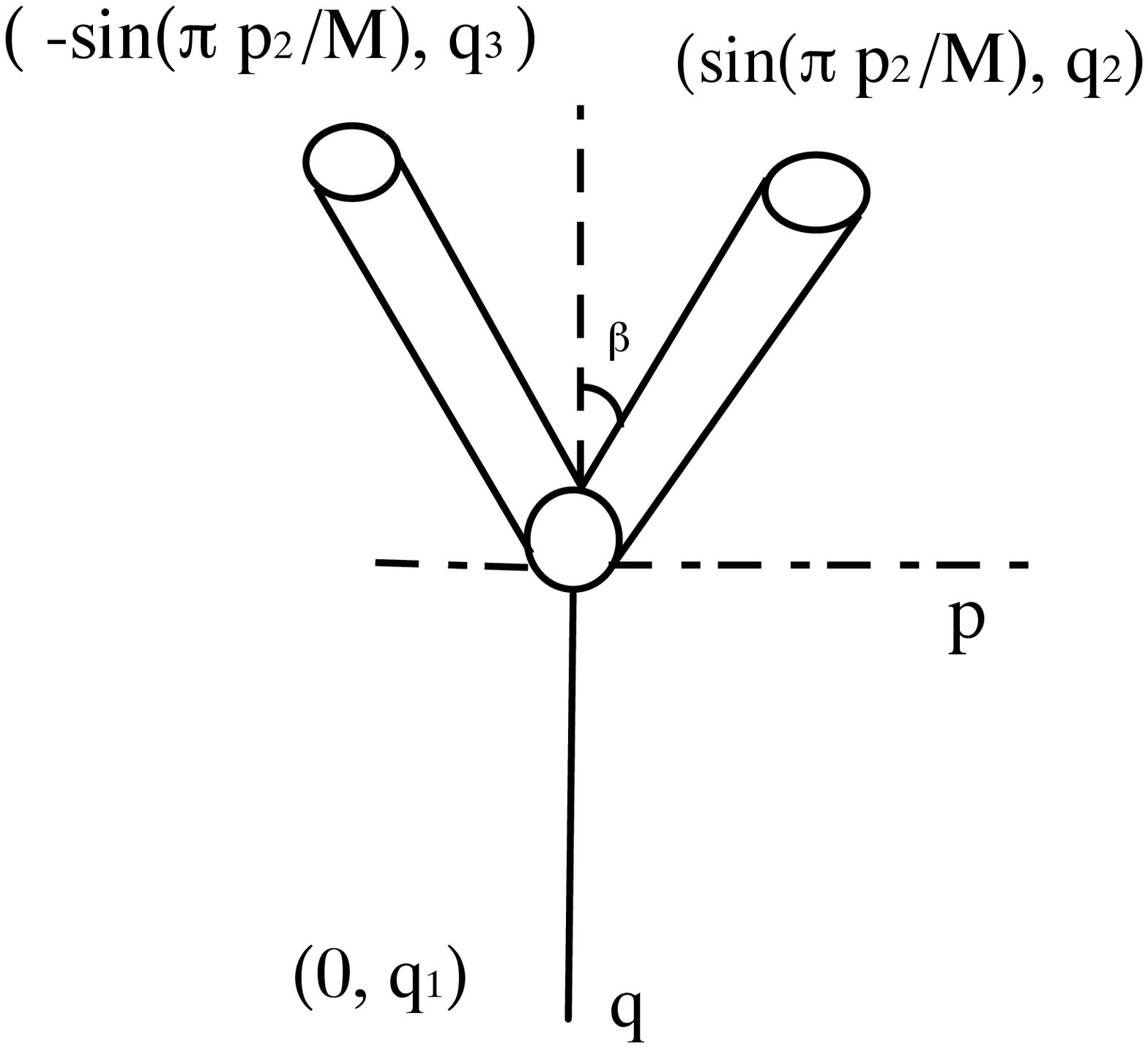}
   \hspace{.8cm} 
   \vspace{-3cm}
\caption{A schematic picture of the three-string junction. The figure on the left shows the junction
of $(p,q)$-strings as observed effectively. The figure in the middle shows a different interpretation of the junction where each string is a wrapped D3-brane with appropriate electric and
magnetic fluxes.  The figure
on the right corresponds to our three-string junction in the KS throat where one string should be in the form of a $(0,q)$-string as required by Eq(\ref{const}).
\label{string}
}
\end{figure}

\subsection{Construction of the junction in the throat}
Having laid down the background and the construction of a $(p,q)$-string in the throat, we are ready to formulate the three-string junction.
As mentioned before, each $(p,q)$-string is a wrapped D3-brane with appropriate units of electric and magnetic fluxes. After integrating out the angular parts of the D3-brane action, each string effectively looks like a one-dimensional object
as in Eq(\ref{action2}). 
%Recently the three-string junction in a flat background was studied in \cite{Copeland:2006if}. We will borrow this formalism for our case of a $(p,q)$-string junction.

We assume that the three strings are coplanar in the $(X^1,X^2)$-plane
and that they are semi-infinite and meet at a point. They form a {\bf Y}-shaped junction as shown in {\bf Fig. 1}.
We are interested only in static junctions, so we do not consider any time-dependent quantities in
the action. It is assumed that the point where the string junction is formed is a stationary point. We derive the conditions under which this can be satisfied.
Each string's worldsheet is equipped with a $(\tau, \sigma)$ coordinate system. 
We may choose $\tau=X^0$, where $X^0$ is the space-time time coordinate, 
and $-\infty < \sigma < \infty$. The strings' spatial orientations in the $(X^1, X^2)$ plane are labelled by $X_i^s$, where $ i=1,2,3$ and $s=1,2$. It should be understood that $i$ labels the string itself while $s$
labels the spatial coordinates of each string in this two-dimensional target plane. It is assumed that the strings do not perturb the flat background metric and fields.

We start with the action of the D3-brane given in Eq(\ref{action1}). After integrating out the angular parts, it gives us an action similar to Eq(\ref{action2}) for a $(p,q)$-string. We use the coordinate $\sigma$ to parameterize the spatial 
extension of each $(p,q)$-string. This coresponds to $h^4 \rightarrow h^4 X_i'^2$
in Eq(\ref{action2}),
where $'$ indicates differentiation with respect to $\sigma$.
Following the formalism of \cite{Copeland:2006if}, the action of the system of three $(p,q)$-strings to form a {\bf Y}-shaped junction is 
\ba
\label{S1}
S= \sum_i \int  d \tau \, d \sigma \,
 \left(-\Delta_i \sqrt{ h^4 X_i'^2 - \la^2 { F_{0\sigma}^i} ^2} + \Omega_i\, F_{0\sigma}^i  \right)
\theta(-\sigma)
  \nonumber\\
+ \int d \tau\, \left[  f_i . (X_i-\bar X) + g_i (A_i  -\bar A)  \right].
\ea
The $\theta$ function 
implies that the strings are extended for $-\infty < \sigma <0$, meet at $\sigma=0$ and vanish at 
$X=\bar X$ for $\sigma >0$. The last two terms in the above action contain Lagrange multipliers $f_i$ and $g_i$ which enforce the constraints that the strings meet at $X=\bar X$, with equal gauge field potential 
$A_i=\bar A$. In this notation $F_{i\,  \tau \sigma}= \partial_\tau A_{i\,  \sigma}- \partial_\sigma A_{i\, \tau} $. We choose the gauge where $A _ {i\, \sigma} =0$ and to simplify the notation we set the convention 
$A_{i\,  \tau} \equiv  A_i$. The definitions of $\Omega_i$ and $\Delta_i$ are given by Eqs 
(\ref{Omega}), (\ref{delta}) and (\ref{psi}), replacing $q$ by $q_i$ and $\psi$ by $\psi_i$. 

A schematic view of the junction is given in {\bf Fig. 1}. 
One important issue is how to cover the boundary,
here $S^2$, when the three wrapped D3-branes meet. A procedure to cover the boundaries was used in \cite{Verlinde:2006bc} to glue two D5-branes which share a common boundary. 
In order to take care of the boundary properly,
we assume that the D3-branes wrap the same $S^2$. In other words the $|\psi_i|$, if non-zero, are identical. This choice corresponds to a limited class of three-string junctions. In general one may consider the case when the wrapped D3-branes do not wrap the same $S^2$. This can represent strings ending on point-like charges, see {\bf Fig. 2}. We postpone this case for the next section.   

The equation of motion for the fields $X_i^s$ is
\ba
\label{eom1}
\partial_\sigma \left(  \frac{ h^4\Delta_i {X^s_i}' }{  \sqrt{h^4 {X_i'}^2 - \la^2 {A_i^{\prime} }^2 } } \theta(-\sigma) 
 %+q_i C_{(2)\, \tau s}  \theta(-\sigma) 
\right)= -f_i^s \, \delta(\sigma),
\ea
which yields the following boundary condition at $\sigma=0$:
\ba
\label{bc1}
f_i^s =  \frac{ h^4 \Delta_i {X^s_i}' }{ \sqrt{ h^4{X_i'}^2 - \la^2 {A_i^{\prime} }^2 } }. %- q_i  C_{(2)\, \tau s} 
\ea
Similarly the equation of motion for the $A_i$ is
\ba
\partial_\sigma \left(  \frac{\lambda^2 \Delta_i A_i' }{ \sqrt{ h^4{X_i'}^2 - \la^2 {A_i^{\prime} }^2 } } 
\theta(-\sigma) - \Omega_i \theta(-\sigma)  \right)= g_i \delta(\sigma),
\ea
with the boundary condition
\ba
\label{bc2}
 g_i= -\frac{\Delta_i  A_i' \lambda^2 }{ \sqrt{h^4 {X_i'}^2 - \la^2 {A_i^{\prime} }^2 } } 
 +\Omega_i.
 \ea
Variation with respect to the Lagrange multipliers $f_i$ and $g_i$ imposes the conditions  $X_i=\bar X$ and $A_i=\bar A$ at $\sigma=0$. Variation with respect to $\bar X$ and $\bar A$
respectively implies that 
\ba
\label{fg}
\sum_i f_i= \sum_i g_i = 0\, .
\ea

The electric charge $p_i$ on the worldvolume of each $(p,q)$-string is defined
by $p_i =\delta {\cal L}/ \delta F_{i\, \tau \sigma}$ which gives 
\ba
\label{p}
p_i=g_i =-\frac{\Delta_i  A_i' \lambda^2 }{ \sqrt{h^4 {X_i'}^2 - \la^2 {A_i^{\prime} }^2 } } 
 +\Omega_i \, .
 \ea
Combined with Eq.(\ref{fg}), this yields the electric charge conservation $\sum_i p_i=0$ at the junction.

Similar charge conservation also holds for the $q_i$. To see that suppose we turn on a hypothetical background $C_{(2)}$ field along the $(\tau,\sigma)$ directions. This induces a Chern-Simons term for each
string and we obtain the following extra contribution to the action
\ba
S_{CS}=  \sum_i  q_i\int  d \tau \, d \sigma \,  \partial_\sigma X_i^s  C_{(2)\, \tau s} \theta(-\sigma).
\ea
In the above formula, by a coordinate transformation, the component of $C_{(2)}$ along the $X^s$ direction of the target plane is written.
In ten dimensions, which one may consider as the bulk,  
the theory is invariant under the gauge transformation 
$C_{(2)\, \tau \sigma} \rightarrow C_{(2)\, \tau \sigma} +
 \partial_\sigma \Lambda$, where $\Lambda$ is a scalar. For a string with no boundary, i.e. closed strings or infinite strings, the string action is invariant under this gauge transformation. 
 However, in our model
 the strings terminate at $\sigma=0$, so this is the boundary of the strings. Upon this gauge transformation 
 on $C_{(2)}$, the string actions  get a surface term 
 \ba
\delta S_{CS}= \sum_i q_i \int d\tau d\sigma \partial _\sigma X_i^s \partial_s \Lambda
 = 2  \sum_i q_i \int d\tau \Lambda(\sigma=0)\, .
 \ea
To keep the action invariant under this gauge transformation we therefore require that
$\sum_i q_i=0$, which is the D-string charge conservation as desired.

The Hamiltonian of the system is
\ba
\label{H1}
{\cal H } &=& \sum_i p_i F_{i\, \tau \sigma} - {\cal L}\nonumber\\
&=& \sum_i f_i^s {X^s_i}',
%\sum_i \left( \frac{ g_s^{-1} |q_i|  {X_i'}^2 }{ \sqrt{ {X_i'}^2 - \la^2 {A_i^{\prime} }^2 } } 
%-  q_i \partial_\sigma X_i^s  C_{(2)\, \tau s}  \right)
\ea
where $f_i$ is given in Eq.(\ref{bc1}).

The energy of the system is given by $E=\int d\sigma {\cal H}$ whereas the tension of each string $T_i$ is defined such that $E= \sum_i \int dX^s  T_{i\, s} $. Using the above expression for the Hamiltonian, one obtains 
\ba
\label{Ti}
T_i= f_i= \frac{ h^4 \Delta_i\, {X_i}' }{ \sqrt{ h^4{X_i'}^2 - 
\la^2 {A_i^{\prime} }^2 } }  \, .
\ea
Furthermore, from Eq.(\ref{fg}) we conclude that 
\ba
\label{balance}
\sum_i T_i=0\, .
\ea
Geometrically, this means that the vector
sum of the string tensions vanishes. This indeed is the necessary condition for the string junction point to be stationary \cite{Rey:1997sp}.

%Due to diffeomorphism invariance on the string worldsheet one can choose the gauge such that ${X_i'}^2=1$. 
From Eq (\ref{p}) the gauge field is given by
\ba
\la A_i'= \frac{h^2(p_i- \Omega_i ) | X_i'|}{ \sqrt{ (\la  \Delta_i)^2 +  (p_i- \Omega_i )^2 } }
.
\ea
Define each string's direction by the unit vector $\hat{n}_i=\frac{X_i'}{|X_i'|}$. 
Using the above equation for the gauge field in Eq(\ref{Ti}) for the tension of strings we obtain 
\ba
\label{Ti2}
 T_i 
= \frac{h^2}{\la}\, \sqrt{ \frac{q_i^2}{g_s^2} + \frac{ M^2}{\pi^2}\,
 \sin^2 \left(  \frac{\pi p_i}{M}\right) }\, \hat{n}_i %\frac{X_i'}{|X_i'|}
\ea
which is in agreement with Eq(\ref{E}).

The above results have the following interesting geometrical interpretation. Consider the coordinate system where $p$ and $q$ represent the horizontal and
the vertical axes respectively as in {\bf Fig. 1}.
Then each string's direction in this coordinate system is given by
\ba
\label{ni}
\hat{n}_i= \frac{1}{|T_i|} \left(  \frac{ M}{\pi}\,  \sin  \frac{\pi p_i}{M} \, , \,   q_i \right).
\ea
Defining $\beta_i$ as the angle between the $i$-th string and the $q$-axis, one obtains
\ba
\label{angle}
\tan \beta_i = \frac{ M  }{\pi q_i}\,   \sin \left(  \frac{  \pi p_i}{M} \right)  .
\ea
In the limit where $M\rightarrow \infty$, this reproduces the known result for the flat background
\cite{Dasgupta:1997pu}.

For the junction point to be stationary  
one obtains from Eqs (\ref{balance}), (\ref{Ti2}) and (\ref{ni}) that 
\ba
\sum_i \sin ( \pi p_i/M ) = 0 \quad \quad , \quad \quad \sum_i q_i =0
.
\ea
The second equation is automatically satisfied due to D-string charge conservation. Combining the first equation with the electric charge conservation $\sum_i p_i=0$ implies that 
\ba
\label{const}
\prod_i  \, \sin \left(  \frac{\pi p_i}{M} \right) =0
.
\ea
This means that one of the strings should be in the form of a D-string, i.e. a $(0,q)$-string, while
the remaining two strings are in the form of $(p,q)$-strings with opposite $p$ charges. In other words, the system contains $(0,q_1), (p_2, q_2)$ and $(-p_2,q_3)$
strings. This configuration corresponds to the right hand side figure in {\bf Fig. 1}.

Of course, the above constraints are expected. As mentioned in the second paragraph after Eq (\ref{S1}), in order to cover the boundaries properly the $|\psi_i|$ are made equal, if non-zero. This means that the two D3-branes which represent the $(p_2,q_2)$ and $(-p_2, q_3)$-strings wrap the same $S^2$ inside $S^3$, but with opposite orientations. The D3-brane corresponding to the $(0,q_1)$-string, on the other hand, shrinks to a point on this $S^2$, which we label by $\psi=0$.  

As mentioned, the above construction corresponds to a limited class of three-string junctions.
It would be an interesting exercise to consider the general $(p,q)$-string junction with the boundary issue properly addressed. In some situations this may require that the D3-branes end up on a spherical D3-brane. We study this case briefly in next section.

\section{Cosmic Necklaces}

\begin{figure}[t]
\vspace{-3cm}
%\vspace{1cm}
  % \centering
   \hspace{2.1cm}
   \includegraphics[width=4in]{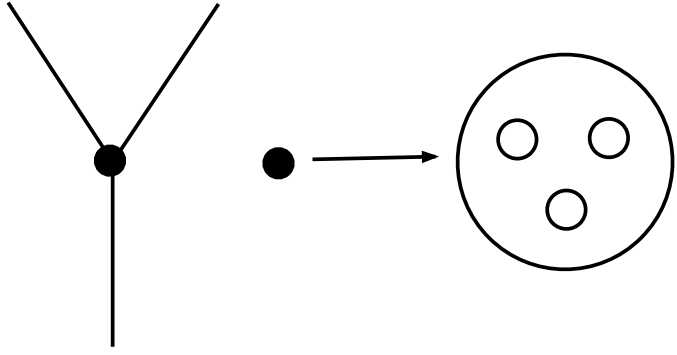} 
   \vspace{-5cm}
\caption{Here a microscopic realization of cosmic necklaces is presented.
The figure on the left hand side represents the junction of three strings on a 
bead or dyon. The figure on the right hand side shows how a dyon is constructed from a wrapped D3-brane. The D3-brane is an $S^3$ with three holes.
Each hole represents an $S^2$ that the D3-branes corresponding to the $(p,q)$-strings can wrap. The total charge of the system is conserved and all boundaries on $S^3$ are covered properly.
\label{bead}
}
\vspace{1cm}
\end{figure}

In this section we present a microscopical realization of a system of
 $(p,q)$-strings which end on point charges, 
i.e. dyons \cite{Hindmarsh:1985xc}. 
The cosmological implications of this system were recently studied in \cite{Leblond:2007tf} and this configuration is dubbed ``cosmic necklaces''
\cite{Berezinsky:1997td}. The total charge of the system
is conserved: $\sum_i p_i=P$ and $\sum_i q_i=Q$ where
$P$ and $Q$ represent the electric and magnetic charges of the dyon or ``bead''.
%In \cite{Leblond:2007tf} it was argued that with some moderate assumptions regarding the populations of lowest tension family of $(p,q)$ strings and their decay channels, the system reaches a scaling regime so it is cosmologically safe. 

To realize this picture microscopically, as in the previous section, we interpret the long $(p,q)$ cosmic strings as wrapped D3-branes with $p$ and $q$ units of electric and magnetic flux respectively. These D3-branes wrap the $S^2$ inside the $S^3$ and have one leg along $x^{\mu}$.
The bead, on the other hand, is a D3-brane with $P$ and $Q$ units
of electric and magnetic flux which wraps the entire $S^3$ at the bottom of the throat. For a four-dimensional observer, this wrapped D3-brane looks like a point-like charged object, as desired. A schematic view of this idea for a three-string junction is presented in {\bf Fig 2}.
In this interpretation the $S^3$ has three holes in its surface with each hole representing a $S^2$. A $(p,q)$-string can wrap around this $S^2$, so all boundaries are covered properly. From a four-dimensional point of view it is nothing but a three-string junction on a dyon.
It will be an interesting exercise to work out the details of this construction such as the tension formula and the conditions for static configuration, as performed in the previous section.
Similarly, one can also construct the configuration where two $(p,q)$-strings end on a bead or a bead and an anti-bead are trapped in a closed $(p,q)$-string loop.

The important quantity in the cosmological evolution of a web of cosmic necklaces as compared to a web of $(p,q)$-strings is the quantity $r$ \cite{Berezinsky:1997td, Leblond:2007tf}
defined by
\ba
r=\frac{M_b}{\mu\, d},
\ea
where $M_b$ is the mass of the bead, $\mu$ is he tension of the cosmic string and 
$d$ is the typical interbead distance along the string.
If $r <<1$ during the network evolution, then one may safely neglect the effect of beads and the web of cosmic necklaces effectively follows the evolution of the web of $(p,q)$ cosmic strings. On the other hand, if $r>>1$, then the web is mostly dominated by beads and cosmologically it is a disaster. Effectively this represents a web of
massive monopoles attached by light strings. It is well-known that massive monopoles will overclose the universe quickly \cite{Preskill:1979zi}. 
In our construction of cosmic necklaces, we notice that the mass of the bead is
given by the tension of D3-brane wrapped around $S^3$
:
\ba
M_b&=& 8 \pi\, h\,  (g_s M \alpha')^{3/2}\,  T_3 \nonumber\\
&=&\frac{M^{3/2}g_s^{1/2}}{\pi^{2} \alpha'^{1/2}}\, h \sim h\, m_s.
\ea
In this expression, $h$ stands for the warp factor at the bottom of the throat.
This just indicates that the physical mass is redshifted inside the warped throat. In the second line $m_s$ represents the scale of string theory
where $m_s \sim \alpha'^{-1/2}$. 
Using the above formula for the mass of the bead, we obtain
\ba
r \sim  (h\, m_s\, d)^{-1}
,
\ea
where the relation $\mu \sim h^{2} m_s^2$ for the cosmic string tension  (from Eq.(\ref{Ti2})) has been used. The quantity $h\, m_s$ represents the physical mass scale of the throat where the junction is formed. It is the same throat 
where brane inflation takes place. 

%The scale of inflation could be anything from
%TeV to GUT. 
We can get useful information on the magnitude of $r$ at the early stage of network evolution.
In brane-antibrane inflation, it is shown that point-like defects, like 
monopoles, are not copiously produced \cite{Sarangi:2002yt}.  
This means that the interbead distance $d$ is typically bigger than the Hubble radius at the end of inflation. Suppose $T_r$ and $H_r^{-1}$ are the temperature of the Universe and the Hubble radius at the end of reheating, respectively. We have $H_r \sim  T_r^2/M_P$, where $M_p$ is the Planck mass. Having $d\ge H_r^{-1}$, as argued above, implies that
\ba
 r \le \left(\frac{T_r}{h\, m_s}\right) \left(\frac{T_r}{M_P}\right).
\ea
In the above equation, the first bracket is less than unity since the reheating temperature is smaller than the scale of inflation $h \, m_s$. The second bracket is also much smaller than unity for $T_r$ not bigger than the GUT scale.
  %Taking $d$ to be comparible to cosmic size at the end of inflation(or at least much bigger than the typical size of cosmic string core or the radius of monopole which is $ \sim m_s^{-1}$)
One can readily see that $r$ 
is many orders of magnitude smaller than unity at early stage of network evolution.
 This is a direct consequence of the facts that (a) in this model the monopoles are not copiously produced and (b) we have only one mass scale: $h\, m_s$. In conventional models when both monopoles and cosmic strings could be present,
the mass of the monopoles and the tension of the cosmic strings have
different origins. This is due to different symmetry breaking which happens at different energy scales in the early Universe. This 
results in different mass scales for monopoles and cosmic strings and in those models $r$ can be bigger than or close to unity.

Although the initial value of $r$ is very small it  may increase in later stage of cosmic evolution. 
This problem was studied in  \cite{Leblond:2007tf}.
It was argued that if the average loop size in the scaling regime is larger than $10^{3} G \mu\, t$ , where
$t$ is the cosmic time and $G$ is the Newton constant, 
then the network of cosmic strings with beads are cosmologically safe. 
 %This indicates that the network of cosmic necklaces will effectively follow the network of cosmic strings with junctions, and for cosmological purposes one may safely neglect the effects of beads.

\section{Semi-local defects in the throat}

In the third and the final case, the D3-branes wrap zero-cycles in the throat, i.e they are points in the 
throat and stretch along the $x^{0,1,2,3}$ directions. 
Here they are useful because we can construct semi-local defects using these branes. semi-local
defects, as discussed in earlier papers, are non-topological defects that are nevertheless stable for a long 
period of time and their existence depends more on the presence of global symmetries than that of local ones\footnote{This 
also means that a sudden emergence of global symmetries may render an unstable defect stable for some time. 
For details on this see \cite{Preskill:1992bf}.}. 
These 
defects were first constructed by Achucarro and Vachaspati \cite{Vachaspati:1991dz}
and have important applications in the fields of 
cosmology, QFT and string theory. 

Our approach to studying
semi-local defects in the throat will be to first determine the possible change in the background 
once branes are incorporated\footnote{Recall that in the previous sections we have ignored the 
backreactions of the branes on the geometry. This is because we wanted to construct ($p,q$)-strings and their 
junctions in our 3+1-dimensional spacetime, so the backreaction effects on the internal geometry would have little
effect and thus could be
ignored. For semi-local strings/defects we need ($p, q$)-strings and junctions {\it inside} the internal space
and therefore their masses and tensions will typically depend on the backreactions. Thus we cannot ignore 
them now.}. 
The branes in question are the seven branes as well as the D3-branes. 
We will be able to divide the internal space into two distinct regions: A and B. In {Region B} we have 
seven branes and in {Region A} we will have the D3-branes. 
These are the
D3-branes that form the third scenario of our set-up, namely, they do not wrap any cycles inside the 
throat geometry. Semi-local defects appear on the D3-branes. 
In fact it is interesting to note that to study semi-local defects with higher global 
symmetries we would necessarily have to incorporate the three-string junctions that we studied earlier 
in the paper. These details will be discussed later in this section. Our first job is to determine the 
metric for the present case. We expect the metric to take the following form
\ba
\label{metform}
ds^2 ~ = ~ F_0~ds^2_{0123} ~ + ~ F_0^{-1} ds^2_{\cal M},
\ea
where $ds^2_{\cal M}$ is the internal six-dimensional metric and $F_0$ is a harmonic form that will be specified
later.
Here we will first elaborate the local scenarios in both the regions A and B, and then try to argue for the 
existence of these defects using various background constraints.\\
 
 Readers who are interested in the final result 
can skip section 4.1 and go directly to section 4.2 where the main constructions and their consequences are presented.

\begin{figure}[t]
\vspace{-5cm}
  % \centering
   \hspace{2.1cm}
   \includegraphics[width=4in]{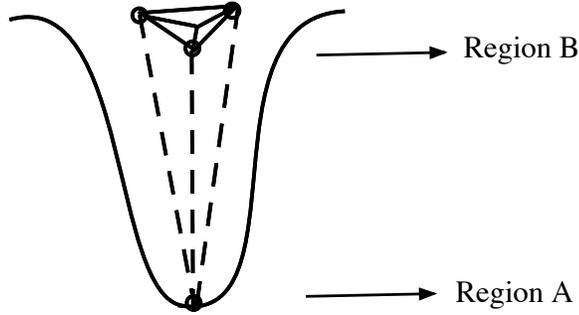} 
   \vspace{-3cm}
\caption{A schematic view of a semi-local defect is shown in the throat. At the bottom of the throat, which we 
called region A in the text, we have D3-branes that are connected by a network of ($p, q$)-strings to the 
seven branes, located  
at the top of the 
throat. We call this region B in the text. These seven branes are connected by 3-string junctions as well as the 
($p,q$)-strings. The complete network of branes, junctions and strings is responsible for creating semi-local defects
on the D3-branes.
\label{semiString}
}
%\vspace{1cm}
\end{figure}

\subsection{The local metrics in two regions}

The local story is however not as simple as that encountered earlier in \cite{Dasgupta:2006yd} as it is 
complicated by the presence of a three-cycle at the end of the throat. This three-cycle is similar to the 
three-cycle present at the end of the throat in Klebanov-Strassler type models or the geometric transition 
models. As will be clear soon, our background is a close approximation to both these models, but differs from them
because of the presence of extra fluxes and branes. We will also consider the backreactions on the geometry 
once the effects of branes and fluxes have been taken into account.  
The resulting metric will be particularly involved as we need to carefully interpolate the metric from one 
region of interest to another. Since the procedure is elaborate, we will go in small steps. The first step is 
clear:
we can argue that 
far away from the three-cycle the local geometry can be given by
\ba
\label{locmet} 
ds_{\rm local}^2 ~ = ~ {\cal A}_1~ dr^2 + {\cal A}_2~(dz ~+~ f_1~ dx ~+~ f_2~ dy)^2 + 
({\cal A}_3 ~ dx^2 + {\cal A}_4 ~d\theta_1^2) +
\nonumber\\ 
~~~~~~~~~~~~~~ + ~({\cal A}_5~ dy^2 + {\cal A}_6~d\theta_2^2).
\ea
This is almost like a conifold geometry except that the coefficients are not constant and ($r, x, y, z, \theta_{1,2}$) 
are local coordinates with $dz$ being the local $U(1)$ fibration\footnote{Choosing the local coordinates is easy. 
We can go to a patch where the geometry looks like a conifold. The non-compact global metric of this is 
specified by coordinates ($\tilde r, \phi_{1,2}, \psi, \theta_{1,2}$) 
with ${\tilde r}$ the radial coordinate and the others angular
(see the first reference of \cite{Candelas:1989js} for details). 
We specify a point $P_0 = $ ($r_0, \phi_{1,2}^0, \psi_0, \theta_{1,2}^0$), and the local coordinates 
($r, x, y, z, \theta_{1,2}$) measure the 
deviations from $P_0$.}.
In fact with constant coefficients we will not be able to satisfy the equations of motion. The
definitions of these coefficients are exactly the same as in Eq. (2.28) and Eq. (2.29) of \cite{Dasgupta:2006yd},
namely:
\ba
\label{adefex}
{\cal A}_i ~ = ~ 1 ~ + ~ {\alpha_i \over {\cal F}_i(r_0)}r ~ + ~  {\beta_i \over {\cal F}_i(r_0)} r^2,
\ea
\noindent with the subscript $i$ running from $i = 1, 2, 3, 5$ . For $i = 4, 6$ we need to replace Eq. (\ref{adefex})
by ${\cal A}_i \to {{\cal A}_i \over 1 + g(\Theta_i)}$ where $\Theta_{4,6} \equiv \theta_{1,2}$. 
As discussed in 
\cite{Dasgupta:2006yd} the function $g(\Theta_i)$ is another perturbative series in $\Theta_i$ whose details will 
not concern us here. 
The geometry also has two tori given by 
\ba
\label{tori}
dz_1 ~ = ~ dx ~ + ~ i d\theta_1, ~~~~~~~ dz_2 ~ = ~ dy ~ + ~ i d\theta_2
.
\ea
The reason the local geometry has non-trivial coefficients is not just 
because of equations of motion, but because with non-trivial 
coefficients (a) we will be able to accomodate extra terms in the metric without violating equations of motion, and (b) 
it will be easier to add three and seven branes.
So the first 
question is what are the allowed extra terms that we could add to Eq. (\ref{locmet}) if we wanted to specify the 
metric close to the three-cycle?

The most generic terms that we could add to our local metric so as to describe the 
local geometry near the three-cycle are the following:
\ba
\label{addterms}
ds^2_1 ~ = ~ a_1~d\theta_1 d\theta_2 ~ + ~ a_2~dx dy ~ + ~ a_3~dx d\theta_1 ~ + 
\nonumber\\
~~~~~~~~~~~~~~~ + ~ a_4~dx d\theta_2 ~ + ~ a_5~dy d\theta_1 ~ + ~ a_6~dy d\theta_2,
\ea
with $a_i$ coefficients that could in general be functions of all the coordinates. The above terms are very generic but
we can simplify this a little bit. It is easy to see that the following terms of the above metric 
\ba \label{twoterms}
a_3~dx d\theta_1 ~ + ~ a_6~ dy d\theta_2
\ea
can be easily absorbed in $ds^2_{\rm local}$ by changing the complex structure of the two tori Eq. (\ref{tori}). This
will leave only four coefficients in Eq. (\ref{addterms}). It is also interesting to note that the combined metric 
\ba \label{defmet} ds^2_{\cal M} ~ = ~ ds^2_{\rm local} ~ + ~ ds^2_1 \ea can be simplified further if we allow the 
following relations between some of the unknown coefficients: 
\ba \label{afrelation} a_1 ~ = ~ - a_2, ~~~ 
a_4 ~ =~ a_5, ~~~ f_1 ~ = f_1(\theta_1), ~~~ f_2 ~ = ~ f_2(\theta_2). \ea
The above condition can be motivated from various earlier scenarios. The original Klebanov-Strassler model 
\cite{Klebanov:2000hb} has a similar structure at the far IR of the geometry, but differs from our case by the 
absence of three and seven branes. The geometry that is closest to our set-up is the one that was studied in 
\cite{Becker:2004qh} (for short reviews see \cite{Becker:2004ii}). We will soon provide a more rigorous 
way to analyse these coefficients (which was not attempted before). But before that observe that imposing 
the relation Eq. (\ref{afrelation}) on the metric Eq. (\ref{defmet}) gives a result which simplifies quite a bit under a rotation of 
($dx, d\theta_1$) by a constant angle $\alpha$ to:
\ba 
\label{sime} ds^2_1 ~~~ {}^{~\alpha}_{\longrightarrow} ~~~ g~ (d\theta_1 d\theta_2 ~ - ~ dx dy),
\ea
when $a_1$ is made proportional to ${\rm cos}~\alpha$ and $a_4$ is made proportional to ${\rm sin}~\alpha$. 
(The coefficient $g$ will be made more precise later.)
Such 
an identification will allow us to write the background equations in a simple form, as we will see 
below\footnote{This is a little subtle. The transformation 
that we are doing here on $ds^2_1$ Eq. (\ref{addterms}) using Eq. (\ref{afrelation}) is: 
\begin{eqnarray}
&&\left( \begin{array}{c} dx \\ d\theta_1 \end{array} \right) \ {\longrightarrow} \
\left(
\begin{array}{cc}
~~{\rm cos}~\alpha  & ~~{\rm sin}~\alpha 
\\
-{\rm sin}~\alpha  & ~~{\rm cos}~\alpha 
\end{array} \right) \left( \begin{array}{c} dx \\ d\theta_1 \end{array} \right)
\end{eqnarray}
\noindent which doesn't quite keep the $dz$ $U(1)$ fibration invariant. This is not {\it a priori} a problem for 
our case because although we will use this simplification to get the metric, we will always work with the 
full three-cycle metric Eq. (\ref{defmet}).}.

To analyse the background equations of motion, we will use the same technique that was developed in 
\cite{Dasgupta:2006yd}. The method involves an order-by-order expansion in terms of coefficients used to 
write the metric Eq. (\ref{locmet}). We would urge the reader to go through the analysis in 
\cite{Dasgupta:2006yd} as we will not elaborate on the techniques again here. Instead only final results will be 
presented. 

The first few relations are easy to derive as they follow exactly the same procedure studied in \cite{Dasgupta:2006yd}. 
The simplest relations between the coefficients are: 
\ba 
\label{firstrel}
&&{\alpha_1 \over {\cal F}_1} ~ - ~ {\alpha_3 \over {\cal F}_3} ~ - ~ {\alpha_5 \over {\cal F}_5} ~ = ~ 0, 
\nonumber\\
&&{\beta_1 \over {\cal F}_1} ~ -~ {\beta_3 \over {\cal F}_3} ~ -~ {\beta_5 \over {\cal F}_5} ~ = ~ 
{\alpha_3\alpha_5 \over {\cal F}_3{\cal F}_5}.
\ea
These are somewhat similar to the equations that were derived for a resolved conifold background in a geometric
transition setting. This might seem a little puzzling as both the geometry and the topology of our 
present background is very different from the resolved conifold studied in \cite{Dasgupta:2006yd} and 
\cite{Becker:2004qh}. The reason we still have the same relations is that these relations are derived 
from analysing the non-compact radial geometry of our system. Since the radial behavior of our geometry 
cannot be too different from the radial behavior studied earlier, the relations Eq. (\ref{firstrel}) remain 
unaltered. 

The next question therefore is to ask whether the $U(1)$ $dz$-fibration would change or not. Here we can take a cue from
the Klebanov-Strassler solution \cite{Klebanov:2000hb} and the geometrical transition set-up of \cite{Becker:2004qh}.
The $U(1)$ fibrations in both set-ups remain the same. This consideration therefore gives us the following 
relations between the other coefficients:
\ba
\label{secondrel}
&&{\alpha_2 \over {\cal F}_2} ~ + ~ {\alpha_4 \over {\cal F}_4} ~ + ~ {\alpha_6 \over {\cal F}_6} ~ = ~ 0, 
\nonumber\\
&&{\beta_2 \over {\cal F}_2} ~ + ~ {\beta_4 \over {\cal F}_4} ~ + ~ {\beta_6 \over {\cal F}_6} ~ = ~ 
{\alpha^2_4 \over {\cal F}^2_4} ~ + ~ {\alpha^2_6 \over {\cal F}^2_6} ~ + ~
{\alpha_4\alpha_6 \over {\cal F}_4{\cal F}_6},
\ea
which differ from Eq. (\ref{firstrel}) not only in relative signs but also in the presence of extra terms. 
These relative signs are crucial because they tell us that the {global} structure of the manifold will 
{\it not} have a trivial $U(1)$. Furthermore for the simplest scenario with local square tori, there are 
further approximate simplifications:
\ba
\label{simplifications}
&& {\alpha_3 \over {\cal F}_3} ~\approx ~ {\alpha_5 \over {\cal F}_5}, ~~~~~~~ {\alpha_4 \over {\cal F}_4} ~ 
\approx ~ {\alpha_6 \over {\cal F}_6}, 
\nonumber\\
&& {\beta_3 \over {\cal F}_3}~\approx ~ {\beta_5 \over {\cal F}_5}, ~~~~~~~ {\beta_4 \over {\cal F}_4} ~ 
\approx ~ {\beta_6 \over {\cal F}_6}, 
\ea
which are approximate because in the geometric transition set-up the base tori do not seem to have the same 
complex structures, although on the other hand in the Klebanov-Strassler case the equality is exact. Our scenario
lies halfway between these two pictures, and therefore we do not expect  exact equality in 
Eq. (\ref{simplifications}). 

\subsubsection{Gluing in a three-cycle in region A}

Up to now our considerations have followed somewhat closely the ideas developed in \cite{Dasgupta:2006yd}. These
were related to the local metric $ds^2_{\rm local}$ in Eq. (\ref{locmet}). Now we should consider the additional 
corrections in $ds^2_1$. These corrections could succinctly be written as
\ba
\label{dsone}
ds^2_1 ~ = ~ {\cal G}~(d\theta_1 d\theta_2 ~ - ~ dx dy) ~ + ~ {\cal H} ~(dx d\theta_2 ~ + ~ dy d\theta_1)
\ea
where the relative signs have already been explained. Observe that ${\cal G}$ is proportional to $g$ in 
Eq. (\ref{sime}).
The coefficients ${\cal G}$ and ${\cal H}$ have the following
perturbative expansions:
\ba
\label{gh}
&&{\cal G} ~ = ~ 1 ~ + ~ {\alpha_7 \over {\cal F}_7(r_0)}r ~ + ~  {\beta_7 \over {\cal F}_7(r_0)} r^2,
\nonumber\\
&&{\cal H} ~ = ~ 1 ~ + ~ {\alpha_8 \over {\cal F}_8(r_0)}r ~ + ~  {\beta_8 \over {\cal F}_8(r_0)} r^2,
\ea
where $r_0$, as mentioned before, is a point in the manifold around which the local geometry is determined. 
In fact when $r_0$ is 
close to the three-cycle, then the local metric is precisely Eq. (\ref{defmet}). Far away from the three-cycle 
in the throat, the local metric is Eq. (\ref{locmet}) as we discussed before. In addition to that, the 
local consideration also assumes that we can ignore ${\cal O}(r^3)$ terms in Eq. (\ref{gh}). 

The next question would be to ask how the three-cycle Eq. (\ref{dsone}) is glued to the local metric 
Eq. (\ref{locmet}). This is not too difficult to work out:
the background equations will now tell us that the coefficients appearing in Eq. (\ref{gh}) above are 
not independent and can be connected to the other coefficients studied earlier. The relations are somewhat 
similar to Eq. (\ref{secondrel}) but differ in coefficients and relative signs:
\ba
\label{thirdrel}
&&s_1 \cdot{\alpha_7 \over {\cal F}_7} ~ - ~ {\alpha_3 \over {\cal F}_3} ~ - ~ {\alpha_5 \over {\cal F}_5} ~ = ~ 0, 
\nonumber\\ 
&&s_1 \cdot{\beta_7 \over {\cal F}_7} ~ - ~ {\beta_3 \over {\cal F}_3} ~ - ~ {\beta_5 \over {\cal F}_5} ~ = ~ 
{\alpha_3\alpha_5 \over 2~{\cal F}_3{\cal F}_5} ~ - ~
{\alpha^2_3 \over 4~{\cal F}^2_3} ~ + ~ {\alpha^2_5 \over 4~{\cal F}^2_5}, 
\ea
where $s_1$ is a constant {\it tunable} factor that will be determined below. The other coefficients also 
have similar relations in terms of our earlier coefficients:
\ba
\label{fourthrel}
&&s_2 \cdot{\alpha_8 \over {\cal F}_8} ~ - ~ {\alpha_4 \over {\cal F}_4} ~ - ~ {\alpha_6 \over {\cal F}_6} ~ = ~ 0, 
\nonumber\\ 
&&s_2 \cdot{\beta_8 \over {\cal F}_8} ~ - ~ {\beta_4 \over {\cal F}_4} ~ - ~ {\beta_6 \over {\cal F}_6} ~ = ~ 
{\alpha_4\alpha_6 \over 2~{\cal F}_4{\cal F}_6} ~ - ~
{\alpha^2_4 \over 4~{\cal F}^2_4} ~ + ~ {\alpha^2_6 \over 4~{\cal F}^2_6},
\ea
but now 
with a different constant $s_2$ compared to Eq. (\ref{thirdrel}). 
Now using the approximate equalities between the earlier coefficients 
Eq. (\ref{simplifications}) we can easily argue the following equalities (again approximate)
\ba
\label{secapp}
{\alpha_8 \over {\cal F}_8} ~ \approx~ {s_1\over s_2}\cdot{\alpha_7 \over {\cal F}_7} ~~~   {\rm and} ~~~  
{\beta_8 \over {\cal F}_8} ~\approx ~ {s_1\over s_2}\cdot {\beta_7 \over {\cal F}_7}
\ea
where we note that only the ratio ${s_1\over s_2}$ shows up. This ratio is in fact a constant and not tunable, and
one can identify this with the angular transformation $\alpha$ in Eq. (\ref{sime}). Calling this ratio 
${s_2 \over s_1} = {\rm tan}~\alpha$, we can show that 
\ba
\label{sost}
s_1 ~ = ~ {2~{\rm sec}~\alpha~ \alpha_3 \over \sqrt{\alpha_3^2 ~ - ~ {\cal F}^2_3}}, ~~~~~~~ 
s_2 ~ = ~ {2~{\rm cosec}~\alpha~ \alpha_5 \over \sqrt{\alpha_5^2 ~ - ~ {\cal F}^2_5}},
\ea 
where again we can use Eq. (\ref{simplifications}) to simplify the above relations. Comparing now 
Eqs. (\ref{thirdrel}), (\ref{fourthrel}) and (\ref{sost}) we see that the local metric Eq. (\ref{locmet}) is the 
limit of Eq. (\ref{defmet}) when 
\ba\label{slimit} s_1 ~ \to~ \infty, ~~~~~ s_2 ~ \to ~ \infty, \ea
which takes us away from the three-cycle at the bottom of the throat. In this limit however our perturbative 
expansions of the coefficients break down and therefore we cannot impose them on our geometry. It also means that the
three-cycle is located at the far IR of the geometry (i.e deep down the throat of our geometry). 

Solving the above set of relations is simple once we specify the metric of one torus, say $dz_1$. In other 
words once ${\cal A}_3$ and ${\cal A}_4$ are both specified, all other ${\cal A}_i$ can be easily determined. The 
issue however is whether there could be a further simplification by making the $dz_1$ torus a square torus. In general, 
as it turns out, this is not always possible. In our case it depends on whether the angle $\alpha$ is a good 
symmetry or not when extra D3-branes are introduced in the geometry. As mentioned in Eq. (\ref{metform}), putting in 
D3-branes alters the background metric by an overall warp factor $F_0$. Secondly, in this region the 
effects of seven branes are negligible and therefore the geometry is close to the Klebanov-Strassler case. Therefore 
all these considerations suggest   
\ba \label{holu} {\cal A}_3 ~ \approx ~ {\cal A}_4 \ea
\noindent and we have square tori\footnote{Recall that this is not the case for the geometric transition set-up
\cite{Becker:2004qh}.}. With this consideration, the only unknown in our relations would be ${\cal A}_3$ or
${\alpha_3 \over {\cal F}_3}$ and ${\beta_3 \over {\cal F}_3}$, whose values can be determined directly from 
the harmonic relations\footnote{A special case was worked out completely in \cite{Dasgupta:2006yd}.}. For our case 
let us assume that
\ba
\label{athree}
{\cal A}_3 ~ = ~ 1 ~ + ~ {\alpha_3 \over {\cal F}_3(r_0)}r ~ + ~  {\beta_3 \over {\cal F}_3(r_0)} r^2 ~ \equiv ~ 
1 ~ + ~ a_0 ~r ~ + ~ b_0 ~r^2,
\ea
\noindent with $a_0, b_0$ constants. Once we know these coefficients, all other ${\cal A}_i$ can be 
easily determined by solving the set of equations given above. They are given by:
\ba
\label{airesults}
&&{\cal A}_1 ~ = ~ 1 ~ + ~ 2 a_0~r ~ + ~ (a_0^2 + 2b_0)~r^2, ~~~~~~ 
{\cal A}_2 ~ = ~ 1 ~ - ~ 2 a_0~r ~ + ~ (3a_0^2 - 2b_0)~r^2,
\nonumber\\
&&{\cal A}_5 ~ \approx ~ {\cal A}_6 ~ \approx ~{\cal A}_4 ~ \approx ~
{\cal A}_3 ~ = ~ 1 ~ + ~ a_0~r ~ + ~ b_0~r^2
,
\nonumber\\
&&{\cal A}_7 ~ = ~ 1 ~ + ~ {2 a_0\over s_1}~r ~ + ~ {a_0^2 + 4b_0 \over 2 s_1}~r^2 , ~~~~~ 
{\cal A}_8 ~ = ~ 1 ~ + ~ {2 a_0\over s_2}~r ~ + ~ {a_0^2 + 4b_0 \over 2 s_2}~r^2. 
\ea
\noindent Using these therefore the metric in region A can be completely specified once we know the other two 
coefficients $s_1$ and $s_2$. For our case they are given by
\ba
\label{sder}
s_1 ~ = ~ 2 ~{\rm sec}~\alpha~{a_0 \over \sqrt{a_0^2 -1}}, ~~~~~ 
s_2 ~ = ~ 2 ~{\rm cosec}~\alpha~{a_0 \over \sqrt{a_0^2 -1}},
\ea
\noindent with $\alpha$ being the angle by which we made a transformation in Eq. (\ref{sime}).

\subsubsection{The metric in region B}

Although using the limit Eq. (\ref{slimit}) doesn't seem to give us the metric at 
large $r$, we can nevertheless
determine the precise local geometry using a different technique. This technique was used before in 
\cite{Dasgupta:2006sg} to determine the D5-D7 bound-state metric in a conifold (or resolved conifold) background.

However before we go into determining the metric far away from the three-cycle, let us analyse the situation first.
We divided our background into two regions of interest. {Region A} is close to the three-cycle with the local
metric being given by Eq. (\ref{defmet}). {Region B} is far away from the three cycle and is in the limit 
Eq. (\ref{slimit}) where we will put our seven branes. The D3-branes $-$ 
parallel to the seven branes and stretched along $x^{0, 1, 2, 3}$ $-$
will be kept in region A such that they are 
far away from the seven branes. The seven branes wrap a four-cycle along the directions ($z, y, \theta_2, \theta_1$) 
and therefore 
can be separated along the radial direction $r$ and angular direction $x$ (which is related to $\phi_1$ in our 
local coordinates). 
The precise metric is given by
\ba
\label{seveng}
ds^2_{\rm IIB} &~ = ~ & {ds^2_{0123}\over \sqrt{h}} + {1\over \sqrt{h}}\big[d\theta_2^2 ~+~ 
\bar\kappa_2^{-1} dy^2\big] ~ +~ \sqrt{h}\big[ dr^2
 ~+~ dx_-^2\big] + {\kappa \sqrt{h}\over \tilde h}~d\theta_1^2 ~ +
\nonumber\\
&& ~~~~~~~~~~~~~~~~~ +~ {\sqrt{h} \bar\kappa_1 \over \tilde h Q_-}~{\rm cos}^2~\sigma_2 
(dz ~ +~ {f_1}~dx ~+~ {f_2}~dy)^2,
\ea
which is almost of the form Eq. (\ref{locmet}) as one would have expected except that in place of $dx^2$ we
have a modified coordinate $dx_-^2$ that captures the geometry\footnote{In fact the $f_i$ values in the $U(1)$ 
fibration would be slightly different. They are given by 
${f_1} = \sqrt{h} \kappa_1^{-1} ~{\rm tan}~\sigma_1~{\rm sec}~\sigma_2$, 
and ${f_2} = -\kappa_1^{-1} \kappa_2^{-1}~{\rm tan}~\sigma_2$. Only for small $\sigma_i$ are both the $f_i$
proportional to ${\rm tan}~\sigma_i$.}. 
This coordinate is related to $x$ in the following way:
\ba
\label{xdef}
x ~\equiv ~\int dx_- ~{\sqrt{\kappa}~{\rm cos}^2~\sigma_1\over \sqrt{\bar\kappa_1}}~ +~ 
x_+~{{\rm sin}~2\sigma_1 \over 2 {\rm sin}~\sigma_2},
\ea
which also means that the harmonic function dependence for the seven brane is not the naively expected 
result. Rather what we have for $h$ in Eq. (\ref{seveng}) is 
\ba
\label{harmonic}
h ~ = ~ \sqrt{1 ~ - ~ {n\over 4\pi}\sqrt{e^{-\phi_0}~ + ~ \chi_0^2~ e^{\phi_0}}~~{\rm log}~(r^2~+~x_-^2)},
\ea
where $\chi_0, \phi_0$ are the asymptotic values of the axion-dilaton appearing due to the seven branes and $n$ is the
number of coincident seven branes in our set-up\footnote{There are also other non-trivial fields switched on for this 
case in addition to the axion-dilaton. However we expect to see negligible effects of these fields at the tip of 
the throat where D3 branes are located.}. 
As one would expect, this will give rise to $SU(n)$ global 
symmetry on the D3 branes at the end of the throat. The constant angles $\sigma_{1,2}$ are related to the background 
fluxes as well as the $U(1)$ fibration in the geometry Eq. (\ref{seveng}). All other variables appearing in 
the above equations are defined in \cite{Dasgupta:2006sg} to which the reader can refer for more details. 

With these considerations, we are ready to study semi-local defects on the D3-branes. Putting D3-branes in region A is
easy: we already gave the relevant form of the local metric there as Eq. (\ref{metform}) with $ds^2_{\cal M}$ defined
in Eq. (\ref{defmet}). The metric in region B with seven branes is Eq. (\ref{seveng}) above. The regions are 
far apart so as to satisfy the criteria laid down in \cite{Dasgupta:2004dw}; namely, the gauge symmetries on the 
seven branes appear as global symmetries on the D3-branes. 

\subsection{Construction of defects}

Now that we have laid down the metrics in the two regions, it is time to construct semi-local defects. In fact 
construction of semi-local defects relies on three constraints that we illustrate below. These constraints need to
be met otherwise no semi-local defects can form.

\vskip.15in

\noindent $\bullet$ The first constraint can be derived directly from the metric information in region B. The 
full global behavior is unrequired. The constraint is that we need the following integral:
\ba
\label{intecon}
{\cal I} ~ = ~ \int_{\Sigma_4} \sqrt{\kappa ~\bar\kappa_1 \over \bar\kappa_2 ~ Q_-}\cdot {{\rm cos}~\sigma_2 \over 
\tilde h} 
\ea
to be maximised. In fact ideally if ${\cal I} ~ \to \infty$ then the defect formation here would match with the 
original construction of \cite{Vachaspati:1991dz}, \cite{Hindmarsh:1991jq}. This is because the gauge fields
from the seven branes couple to the three branes with a coupling constant proportional to ${\cal I}^{-1}$. Thus
for small ${\cal I}$ any semi-local defect would quickly 
dissociate away. We see that it is not so difficult to adjust the parameters appearing in Eq. (\ref{intecon}) 
to maintain a very large value. The metric of the four-cycle $\Sigma_4$ over which we will be performing the 
integral is given by 
\begin{eqnarray}
\label{metmat}
&& g(\Sigma_4) \ = \
\left(
\begin{array}{cccc}
{\kappa \sqrt{h}\over \tilde h}  & 0 & 0 & 0
\\
0 & {1\over \sqrt{h}} & 0 & 0
\\
0 & 0 & {\sqrt{h} ~\bar\kappa_1 ~ {\rm cos}^2~\sigma_2 \over \tilde h~Q_-} & 
{f_2 \sqrt{h} ~\bar\kappa_1 ~ {\rm cos}^2~\sigma_2 \over \tilde h~Q_-}
\\
0 & 0 & {f_2 \sqrt{h} ~\bar\kappa_1 ~ {\rm cos}^2~\sigma_2 \over \tilde h~Q_-} & 
{\cal Q}
\end{array}
\right) \ ,
\end{eqnarray}
where the diagonal elements of this metric correspond to $g_{\theta_1\theta_1}, g_{\theta_2\theta_2}, g_{zz}$ and 
$g_{yy}$ respectively. The element ${\cal Q}$ appearing above is 
\ba
\label{qdeff}
{\cal Q} ~ \equiv ~ g_{yy} ~ = {1\over \bar\kappa_2 \sqrt{h}} ~ + ~ 
{f^2_2 \sqrt{h} ~\bar\kappa_1 ~ {\rm cos}^2~\sigma_2 \over \tilde h~Q_-},
\ea
where $f_2$ takes the value specific for region B (recall that in region A $f_{1,2}$ are different). Modulo this subtlety
all other definitions remain the same as discussed above. 

\vskip.15in

\noindent $\bullet$ The second constraint can also be derived from the metric information that we laid down 
in the previous section, except that now we need the full global behavior. The metric in question is the 
one orthogonal to the four-cycle $\Sigma_4$ discussed above, and therefore is the line element 
associated with ($r, x$) locally. This line element in region $B$ is 
\ba
\label{lelem}
{ds^2_B \over \sqrt{h}} ~ &=& ~ \left(1~ + ~ {\kappa ~{\rm cos}^4~\sigma_1~{\rm cos}^2~\sigma_2~f_1^2 \over
 \tilde h~Q_-}\right)~dx_-^2 ~ + ~ 
{{\rm sin}^2~2\sigma_1 \over 4~{\rm tan}^2~\sigma_2} \cdot {\bar\kappa_1\over \tilde h ~Q_-}~dx_+^2 ~ + 
\nonumber\\
 && ~~~~~~~~~ + ~ {f_1^2 \sqrt{\kappa \bar\kappa_1}\over \tilde h ~Q_-}\cdot {{\rm cos}^2 ~\sigma_1 ~ 
{\rm cos}^2~\sigma_2 ~{\rm sin}~2\sigma_1 \over {\rm sin}~\sigma_2}~dx_- ~dx_+ ~ + ~ dr^2,
\ea
where $x_\pm$ are not independent or orthogonal coordinates. Rather they are some linear decomposition of $x$ 
described above in Eq. (\ref{xdef}). A ($p,q$)-string stretched in this region will have a mass proportional to
\ba
\label{mass}
m_B ~ = ~ \int_{\bf B} T_{p,q} ds_B,
\ea
where $T_{p,q}$ is the ($p,q$)-string tension (the relevance of the ($p,q$)-string will become clear soon). Observe that 
this is a string stretched in the internal space as opposed to the ($p,q$)-string 
stretched along the spacetime directions that were studied in earlier sections. 

The relevant metric in region $A$, on the other hand,  
can be derived using the order-by-order expansion that we did above. As before, the 
line element is along ($r, x$), and is given as
\ba
\label{aline}
ds^2_A ~ = ~ {\cal A}_1 ~dr^2 ~ + ~ {\cal J}~dx^2,
\ea
where ${\cal A}_1$ is the same variable appearing in Eq. (\ref{locmet}) and whose value was determined in
Eq. (\ref{airesults}). Furthermore,
this time $x$ is not split into $x_\pm$. 
The other variable ${\cal J}$ has the following expansion in powers of $r$:
\ba
\label{jexp}
{\cal J} ~ = ~ J_0 ~ + ~ {\alpha_9 \over {\cal F}_9(r_0)}r ~ + ~  {\beta_9 \over {\cal F}_9(r_0)} r^2
\ea
which differs from our earlier perturbative expansions by the appearance of $J_0$ as the first term. The 
background equations will tell us that the coefficients appearing in this equation are not independent, and 
can be related to earlier coefficients in the following way:
\ba
\label{ggila}
&&{\alpha_9 \over {\cal F}_9} ~ = ~  {\alpha_3 \over {\cal F}_3} ~ + ~ \left({\alpha_4 \over {\cal F}_4} ~ + ~ 
{\alpha_6 \over {\cal F}_6}\right)~f_1^2, 
\nonumber\\
&& {\beta_9 \over {\cal F}_9} ~ = ~ {\beta_3 \over {\cal F}_3} ~ + ~ \left({\beta_4 \over {\cal F}_4} ~ + ~ 
{\beta_6 \over {\cal F}_6} ~ + ~ {\alpha_4 \alpha_6 \over {\cal F}_4 {\cal F}_6}\right)~f_1^2,
\ea
which means, using earlier relations, we can uniquely determine ${\cal J}$ to the relevant order. The coefficient 
$J_0$ is slightly bigger than 1, and is given by 
\ba
\label{jzero}
J_0 ~ = ~ 1 ~ + ~ f_1^2,
\ea
with $f_1$ defined in region $A$ (which from Eq (2.29) of \cite{Dasgupta:2006yd} is a power expansion in $\theta_1$,
the local angular coordinate appearing in Eq. (\ref{locmet}))\footnote{It is interesting to note that in some 
cases, studied for example in \cite{Becker:2004qh}, the power series in $\theta_1$ would converge to give us 
${\rm cot}~\theta_1$. This would imply that $J_0 = {\rm cosec}^2~\theta_1$ with $\theta_1 > 0$.}, and the 
relevant line element in region A can now be easily determined because we can solve the set of 
equations (\ref{ggila}) in terms of $a_0$ and $b_0$ to give us: 
\ba
\label{lela}
&&ds^2_A ~ = ~ \left[1 ~+~  2 a_0~r  + (a_0^2 ~+~ 2b_0)~r^2\right] dr^2 ~+~ 
\nonumber\\
&& + (1 + f_1^2)\left[1 ~+~ {1 + 2f_1^2\over 1 + f_1^2} a_0 r + \left\{{f_1^2\over 1 + f_1^2} a_0^2 ~+~ 
\left({1 + 2f_1^2\over 1 + f_1^2}\right)b_0\right\}r^2\right] dx^2.
\ea 
\noindent Therefore the mass of a 
string stretched in this regime will be given by a relation similar to Eq. (\ref{mass}) with $ds_B$ replaced 
by ${ds_A \over F_0}$ where $F_0$ defined in Eq. (\ref{metform}) is the harmonic form associated with coincident D3-branes.  

The question now is to find the mass of the string stretched between regions $A$ and $B$. For that we need the interpolating metric
between Eq. (\ref{locmet}) and Eq. (\ref{seveng}) i.e. from the regime where at the boundary of region $A$ the size
of the three-cycle is almost zero, to the boundary of region $B$ where $dx_- \approx dx$. Clearly the interpolating
metric is now the metric of a warped conifold! This is also exactly the regime where in \cite{Burgess:2006cb} 
the analysis of uplifting was done. The mass of the string stretched between the three and seven branes is therefore
 
\ba
\label{mab}
m ~ = ~ m_A ~ + ~ m_B ~ + ~ m_{AB},
\ea
with $m_{AB}$ computed using the line element for ($dr, dx$) using a warped conifold\footnote{This should be 
compared to the similar situation encountered in the D3/D7 system \cite{Dasgupta:2002ew}, 
\cite{Dasgupta:2004dw}, \cite{Chen:2005ae} where the mass of the string stretched between three and seven branes
could be computed by extending the result of \cite{Sen:1996sk} to the warped case.}. Our constraint therefore is 
that $m$ computed using the metric should be minimised. 

\vskip.15in

\noindent $\bullet$ Our third constraint is that there should exist an F-theory \cite{Vafa:1996xn}
lift of the background. This is not too difficult to show. We know of the following three cases:

\vskip.1in

\noindent (a) For the warped conifold case, the F-theory 
lift has been shown to exist in \cite{GKP}.

\vskip.1in

\noindent (b) For the warped deformed conifold case, the F-theory lift has been shown to exist in \cite{park} by 
compactifying the manifold. 

\vskip.1in

\noindent (c) For the warped resolved conifold case, the F-theory lift has been shown to exist in the third 
reference of \cite{Becker:2004qh}. In fact a conifold transition can be shown to reproduce the F-theory 
lift of the deformed case also. 

\vskip.15in

\noindent We see that for our case then F-theory description can be 
easily constructed. Once an F-theory description is given $-$ satisfying the other two constraints $-$
it is easy to construct semi-local defects because an F-theory picture 
can be used to construct global symmetries in the set-up (a schematic view of the whole set-up is given in 
{\bf Fig. 3}) {\it without} violating the type IIB equations of motion. 
Assuming the global group is ${\cal G}$, the semi-local
defects can be constructed using the following procedures that were laid down in \cite{Chen:2005ae}:

\vskip.15in

\noindent $\bullet$ Given any global group with the corresponding algebra ${\cal G}$, first find the {\it maximal} regular subalgebra ${\cal H}$. Having a
maximal subalgebra would mean that we could ignore the $u(1)$ groups.

\vskip.1in

\noindent $\bullet$ The subalgebra should be expressible in terms of  a product of two smaller subalgebras, namely ${\cal H} = {\cal H}_1 \times {\cal H}_2$.

\vskip.1in

\noindent $\bullet$ One of the ${\cal H}_i$ should form the {\it
local} gauge symmetry on the D3 brane(s). For example let us
assume ${\cal H}_1$ to be the allowed local algebra (or the local
group) on the D3-brane(s).

\vskip.1in

\noindent $\bullet$ The local group\footnote{We use the same notation for the group and its algebra. The distinction between them doesn't affect the analysis below.}
${\cal H}_1$ should have the required homotopy classification $\pi_n({\cal H}^c_1) = \mathbb{Z}$, where ${\cal H}^c_1$ is the coset       
for the group ${\cal H}_1$.

\vskip.1in

Then semi-local defects can form on the D3-brane(s) worldvolume with a global symmetry ${\cal G}$, provided the energetics also allow for this. The generic breaking
pattern of the groups in this case will be:
\ba
\label{grbrpat}
 {\cal G}_g \times ({\cal H}_1)_l ~~ {}^{~\Phi}_{\longrightarrow} ~~ ({\cal H}_2)_g \times ({\cal H}_1)_g,
\ea
\noindent where the subscript $g$ and $l$ refer to global and 
local symmetries respectively. The corresponding coset manifold ${\cal M}_G$,   for which
  every point
$p$   corresponds to   a semi-local defect that cannot be deformed into 
 another    one   with finite energy, is given by
\ba
\label{semmdef}
{\cal M}_G ~ = ~ {{\cal G} \over {\cal H}_1 \times {\cal H}_2}.
\ea
\noindent In fact the above is the most generic construction of semi-local defects and can be shown to accommodate 
all the known semi-local defects constructed so far! It is also not too difficult to show that the ${\cal M}_G$ form
quaternionic K\"ahler manifolds whose classifications have been done in the mathematics literature by 
Wolf and Alekseevskii \cite{wolf}.

\section{Cosmological implications}

Unlike monopoles and domain walls, cosmic strings are cosmologically safe.
Although cosmic strings cannot be the dominant source for the CMB anisotropies, they can contribute as a subdominant source up to 
a
few percent \cite{Bevis:2007gh}. This will put an upper bound on $G\mu$, the dimensionless number corresponding to cosmic string tension. An even stronger bound may come from recent studies on pulsar timing \cite{Jenet:2006sv}, 
which can give the upper bound $ G\mu \le 1.5 \times 10^{-8}$. This is within the bound obtained in warped brane-antibrane inflation models \cite{Firouzjahi:2005dh}.

However, in these studies %studies so far for the bound on $G\mu$ from CMB 
%or gravitational waves, 
it is assumed that the cosmic strings are like gauge field strings with intercommutation probability $P$ equal to unity.
%There have been some interesting studies for the network of cosmic superstrings
 %\cite{Jackson:2004zg, Tye:2005fn, Copeland:2006if, Leblond:2007tf}, but the bi picture of the evolution of cosmic superstrings is still missing.
The evolution of  cosmic string networks depends sensitively on the intercommutation probability. When two gauge field cosmic strings intersect, they exchange partners and the intercommutation probability is unity, i.e. $P=1$. Reducing $P$ would increase the number density of cosmic strings at the
final scaling regime. This also results in a stronger gravitational wave signal
which may be detected in upcoming gravity waves experiments such as LIGO and LISA \cite{Damour:2004kw}.% or in B-mode polarization from CMB observations .

For cosmic superstrings of different types, the situation is quite different. When strings of different types intersect, usually they 
cannot change partners to intercommute. Instead, a three-string junction of
$(p,q)$-strings, as studied in section 2, is formed. Furthermore, for the case when intercommutation takes place, its probability may be significantly less than unity \cite{Jackson:2004zg, Hanany:2005bc}. In \cite{Jackson:2004zg} the intercommutation probabilities for F-F, D-D and F-D collisions in flat background were calculated.
It was shown that for F-strings, the reconnection probability is of order $g_s^2$. For typical valuse of $g_s$, this would lead to $10^{-3} \leq P \leq 1.$ For D-D collisions, depending on models, one finds that $10^{-1} \leq P \leq 1$, while for F-D collisions $P$ may vary from 0 to 1. 

When three-string junctions are present, the evolution of a cosmic string network is complicated \cite{Copeland:2003bj, Tye:2005fn}.
It is possible that the network evolves into a three-dimensional structure and freezes, for example see \cite{Copeland:2003bj} and references therein. In this case it dominates the energy density of the Universe. This may put strong constraints on string formation in brane inflation. Whether or not the cosmic superstring network with junctions reaches the scaling may depend on $g_s$ and the details of compactification. For example, consider the limiting case when $g_s<<1$. In this case D-strings are much heavier than F-strings by a factor of $g_s^{-1}$.
One may effectively ignore the F-string contribution to the network. If the D-strings share more or less the same intercommutation probability as ordinary gauge strings, then the network reaches the scaling regime. Subsequently, the F-strings evolve separately later on.  

In \cite{Copeland:2006if} the kinematical constraint on string intercommutation
for a network with string junctions was studied. It was shown that two different types of string may not necessarily reconnect into a third string. Depending on their relative velocity, their ratio of tensions and the angle of collision, they may pass through each other or even form a locked system as
an {\bf X} configuration.

The evolution of $(p,q)$-cosmic superstrings with the specific tension formula 
Eq (\ref{E}) was recently studied in 
\cite{Leblond:2007tf}. Due to the non-linearity in the tension formula for F-strings, non-coprime $p$ and $q$ strings can combine to form a $(p,q)$ bound-state. This is different from the flat background case with the tension formula Eq (\ref{flat}).
It was argued that the system reaches a scaling regime. Also in \cite{Leblond:2007tf} the network of $(p,q)$-strings ending on point-like charged objects was studied. It was argued that if the average loop size is larger than $10^{3} G \mu\, t$ in the scaling regime, then the network of $(p,q)$-strings with beads are cosmologically safe.

Another interesting feature of cosmic strings with junctions is their novel implications for lensing. A straight cosmic string produces two identical images
of cosmological objects like galaxies. The lensing from strings at a junction, however, is more nontrivial \cite{Shlaer:2005ry}. For example, for a {\bf Y}-shaped string, three identical images of a single object can form. For
strings in a network of junctions such as in \cite{Sen:1997xi}, the lensing is even richer. This may include
multiple identical images along with images where only parts of the object are
present. It is not very probable to locate strings at a junction in the sky, but if such lensing were observed it could provide a spectacular window to the $(p,q)$-string network.   

Semi-local strings on the other hand, have interesting cosmological consequences as discussed in 
\cite{Achucarro:1998ux, Urrestilla:2001dd, Dasgupta:2004dw}. These strings formed at the end of 
inflation are generically short with monopoles stuck at the two ends. Thus two oppositely oriented 
strings\footnote{Changing orientation means we are changing the monopoles to anti-monopoles.} can either 
combine to form a big string, or merge and shrink to zero size. Only when these strings combine to form 
one long (or infinite) string, can they have large-scale density perturbations. However the probability that 
small semi-local strings would combine to form one infinite string is very low, and therefore will have very little
effect on density perturbations. This in turn means that a copious production of cosmic strings, when converted 
to semi-local strings by enhancing the global symmetry (i.e by increasing the number of seven branes), will become
cosmologically safe. In \cite{Achucarro:2005tu} the evolution of a network of semi-local strings at the
stable regime in flat space-time is numerically studied. It is argued that observational signatures of semi-local strings in CMB may resemble those of global texture or monopoles.

\section{Discussion}

In this paper we studied various solitonic or lump solutions in the throat
region of warped six dimensional manifolds.
The lump solutions that we studied in this paper are 
($p,q$)-string junctions, cosmic necklaces,
semi-local strings and generic semi-local defects. All these are classical
solutions and appear in our
four dimensional universe as defects.
We also showed how ($p,q$)-strings and their junctions in the internal
space (i.e the
warped manifold) are important to study semi-local strings and defects.
These solutions, which appear towards the
end of inflation,
have important applications
in cosmology and therefore a detailed study of these solutions may give us
insight into the cosmological evolution
of our universe and string theory itself.

As an interesting exercise, one can study the BPS properties of the 
$(p,q)$-string junctions in KS throat. Motivated by \cite{Dasgupta:1997pu, Rey:1997sp, Callan:1997kz},
we expected that
the three-string junction studied in section 2 to be supersymmetric. We expect the condition that the junction is 
supersymmetric will 
reproduce the constraints obtained in section 2 for a stationary junction. The semi-local strings on the other hand are
supposedly BPS by construction, and has
been rigorously demonstrated in a warped background of the form $K3 \times
T^2/{\mathcal Z}_2$ \cite{Dasgupta:2004dw}.
But for a warped throat of the form that we studied in section 4, this
still needs to be shown.

In addition to studying classical lumps, we need to discuss their quantum
descendents. The various quantum
descendents would be related to small fluctuations of these solutions. An
attempt to study zero mode fluctuations
of three-string junctions in a flat background was done early on in
\cite{Callan:1998sf}.  For our case this
will be subtle: the small fluctuations of these lumps would have to couple
to the small fluctuations of the
background warped geometry. This would then bring out the important
question of the existence of normal
modes in the system, much like the ones that we study when quantising kink
or Sine-Gordon solitions. Other
questions like, how do these quantum descendents influence large scale
density perturbations or do these
quantum effects have any signature on the post inflationary evolution of
our universe, should now be addressed.

Finally, apart from cosmological applications, semi-local strings and
defects have other applications
in the areas of gauge theories, mathematics etc. For example these strings
could be used to classify
quaternionic K\"ahler manifolds and serve as a potential candidate to form
a link between these exotic manifolds and
non-abelian gauge theories with exceptional global symmetries. These and
other details are beyond the scope of this
paper and will be addressed elsewhere.

\vspace{1cm}

\noindent {\bf \large {Acknowledgements}}

\vspace{5mm}

We thank Robert Brandenberger, Jim Cline and Louis Leblond for useful discussions. 
Two of the authors (KD and HF) would also like to thank the Banff center for hospitality during the last summer 
when the project was initiated. 
This work is supported by NSERC grants. RG is also supported by 
a Chalk-Rowles fellowship.

\end{document}